\documentclass{aa}

\usepackage{amssymb}
\usepackage{amsmath}
\usepackage{graphicx}
\usepackage{natbib}
\usepackage{txfonts}
\usepackage{color}

\bibpunct{(}{)}{;}{a}{}{,} 

\begin{document}

\title{New measurements of $\Omega_m$ from gamma--ray bursts}
\titlerunning{Cosmological parameters and GRBs}
\authorrunning{L. Izzo, M. Muccino, E. Zaninoni, et al.}
\author{L. Izzo$^{1,2}$, M. Muccino$^{1,2}$, E. Zaninoni$^{3}$, L. Amati$^{4}$, M. Della Valle$^{5,2}$}
\institute{$^{1}$Dip. di Fisica, Sapienza Universit\'a di Roma, Piazzale Aldo Moro 5, I-00185 Rome, Italy.\\
$^{2}$ICRANet-Pescara, Piazza della Repubblica 10, I-65122 Pescara, Italy.\\
$^{3}$ICRANet-Rio, Centro Brasileiro de Pesquisas Fisicas, Rua Dr. Xavier Sigaud 150, 22290-180, Rio de Janeiro, Brazil.\\ 
$^{4}$INAF, Istituto di Astrofisica Spaziale e Fisica Cosmica, Bologna, Via Gobetti 101, I-40129 Bologna, Italy.\\
$^{5}$INAF-Napoli, Osservatorio Astronomico di Capodimonte, Salita Moiariello 16, I-80131 Napoli, Italy.\\ }

\abstract
{Data from cosmic microwave background radiation (CMB), baryon acoustic oscillations (BAO), and supernovae Ia (SNe-Ia) support a constant dark energy equation of state with $w_0 \sim -1$. Measuring  the evolution of $w$ along the redshift is one of the most demanding challenges for observational cosmology.}
{We  discuss the existence of a close relation for GRBs, named Combo-relation, based on characteristic parameters of GRB phenomenology such as the prompt intrinsic peak energy $E_{p,i}$, the X-ray afterglow, the initial luminosity of the shallow phase $L_0$, the rest-frame duration $\tau$ of the shallow phase, and the index of the late power-law decay $\alpha_X$. We use it to measure $\Omega_m$ and the evolution of the dark energy equation of state. We also propose a new calibration method for the same relation, which reduces the dependence on SNe Ia systematics.}
{We have selected a sample of GRBs with  1) a measured redshift $z$; 2) a determined intrinsic prompt peak energy $E_{p,i}$, and 3) a good coverage of the observed (0.3Ð10) keV afterglow light curves. The fitting technique of the rest-frame (0.3Ð10) keV luminosity light curves represents the core of the Combo-relation. We separate the early steep decay, considered a part of the prompt emission, from the X-ray afterglow additional component. Data with the largest positive residual, identified as flares, are automatically eliminated until the p-value of the fit becomes greater than 0.3.}
{We strongly minimize the dependency of the Combo-GRB calibration on SNe Ia. We also measure a small extra-Poissonian scatter of the Combo-relation, which allows us to infer from GRBs alone $\Omega_M =0.29^{+0.23}_{-0.15}$ (1$\sigma$) for the $\Lambda$CDM cosmological model, and $\Omega_M =0.40^{+0.22}_{-0.16}$, $w_0 = -1.43^{+0.78}_{-0.66}$ for the flat-Universe variable equation of state case.}
{In view of the increasing size of the GRB database, thanks to future missions, the Combo-relation is a promising tool for measuring $\Omega_m$ with an accuracy comparable to that exhibited by SNe Ia, and to investigate the dark energy contribution and evolution up to $z \sim 10$.}


\offprints{\email{luca.izzo@icra.it}\\
\email{muccino@icra.it,elena.zaninoni@gmail.com}}

\date{}

\maketitle

\section{Introduction}\label{sec:1}

Gamma-ray bursts (GRBs) are observed in a wide range of spectroscopic and photometric redshifts, up to $z \sim 9$ \citep{Salvaterra2009,Tanvir2009,Cucchiara2011}. This  suggests that GRBs can be used to probe the high--$z$ universe, in terms of  investigating the re--ionization era, population III stars, the metallicity of the circumburst medium, the faint--end of galaxies luminosity evolution \citep{DElia2007,Robertson2012,MacPherson2013,Trenti2013,Trenti2015}, and using GRBs as cosmological rulers \citep[e.g.][]{Ghirlanda2004,Dai2004,Amati2013}. The latter  works show that GRBs, through the correlation between radiated energy or luminosity and the photon energy at which their $\nu F(\nu)$ spectrum peaks $E_{p,i}$, admittedly with a lower level of accuracy, provide results in agreement with supernovae Ia (SN Ia, \citealp{Perlmutter1998,Perlmutter1999, Schmidt1998,Riess1998}), baryonic acoustic oscillations (BAO, \citealp{WiggleZ}), and cosmic microwave background (CMB) radiation \citep{Planckfirstrelease}. The universe is spatially flat \citep[e.g.][]{deBernardis2000}, and it is dominated by a still unknown vacuum energy, usually called dark energy, which is responsible for the observed acceleration.
Measuring the equation of state (EOS, $\omega = p / \rho$, with $p$ the pressure and $\rho$ the density of the dark energy) is one of the most difficult tasks in observational cosmology today. Current data \citep{Union2.1,Planck2015} suggest that $w_0\sim-1$ and $w_a\sim0$,  the expected values for the cosmological constant. Although these results are probably sufficient to exclude a very rapid evolution of dark energy with $z$, we cannot yet exclude  that it may evolve with time, as originally proposed by \citet{Bronstein1933}. In principle, with SNe Ia we can push our investigation up to $z \approx 1.7$ \citep{Union2.1}. However we note that the possibility of detecting SNe Ia at higher redshifts will depend  on the availability of next-generation telescopes and also on the time delay distribution of SNe Ia  \citep[see Figure 8 in][]{Mannucci2006}.

A solution to this problem may be provided by GRBs: their redshift distribution peaks around $z \sim 2 - 2.5$ \citep{Coward2013} and extends up to the photometric redshift of $z = 9.4$, \citep{Cucchiara2011}. Therefore, given this broad range of $z$ and their very high luminosities, GRBs are a class of objects suitable to exploring the trend of dark energy density with time \citep{Lloyd1999,Ramirez2000,Reichart2001,Norris2000,Amati2002,Ghirlanda2004,Dai2004, Yonetoku2004,Firmani2006,Liang2006,Schaefer2007,Amati2008,CapozzielloIzzo2008,Dainotti2008,Tsutsui2009,Wei2014}.  There are two complications connected with these approaches. First, the correlations are always calibrated by using the entire range of SNe Ia up to $z = 1.7$; therefore, this procedure strongly biases the GRB cosmology and for our purposes we need the highest level  of  independent calibration possible and second,  the data scatter of these correlations is not tight enough to constrain cosmological parameters (in this work we refer to the different extra-scatters published in \citealp{Margutti2013}), even when a large GRB dataset is used. In this work we present a method that can override the latter and minimize the former.

\begin{figure}
\centering
\includegraphics[width=\hsize,clip]{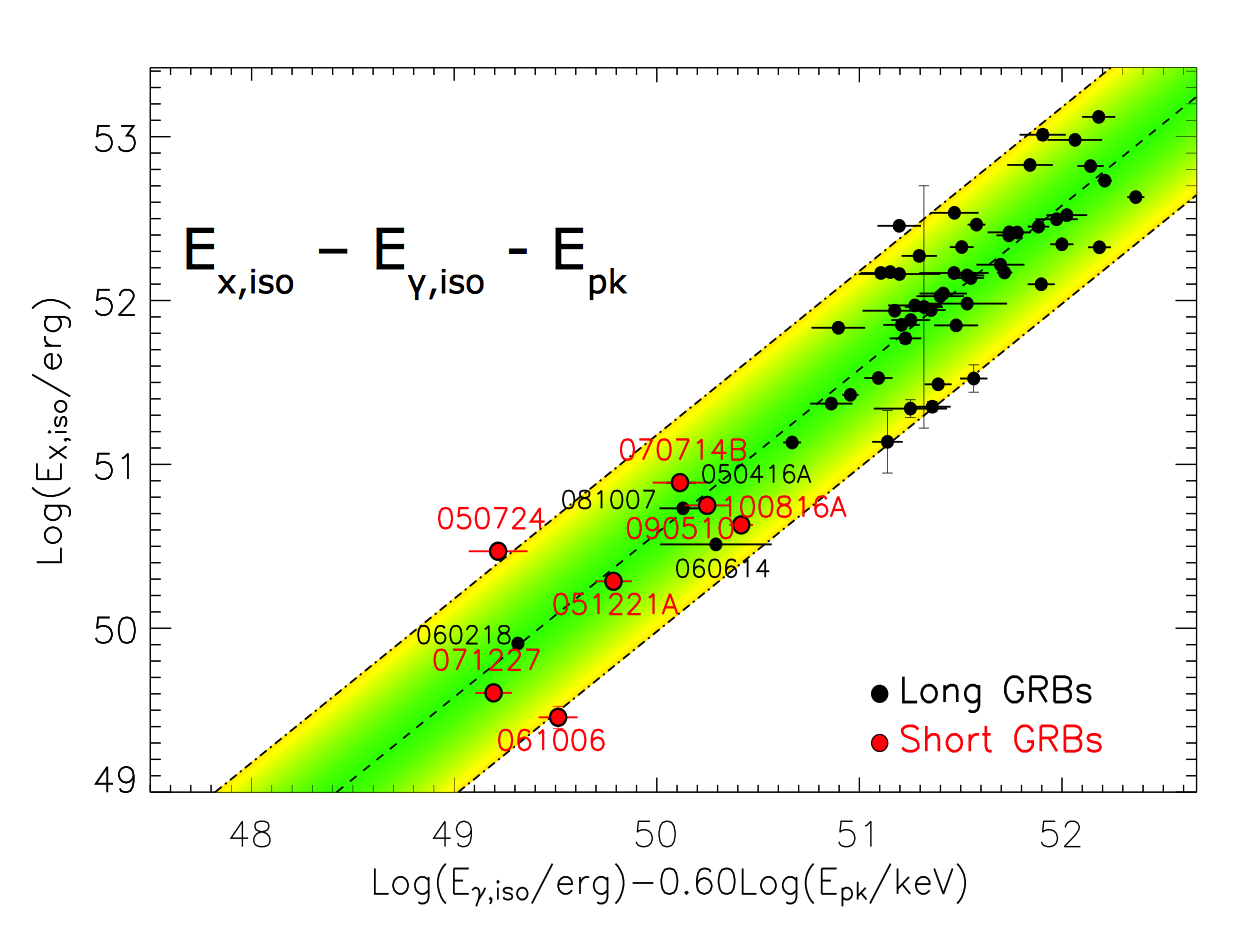}
\caption{The $E_{X,iso}$--$E_{\gamma,iso}$--$E_{p,i}$ correlation proposed by \citet{Bernardini2012} and \citet{Margutti2013} \textit{(courtesy R. Margutti)}.}
\label{fig:noM}
\end{figure}

Very recently \citet{Bernardini2012} and \citet{Margutti2013} (hereafter B12 and M13, respectively) have published an interesting correlation that connects the prompt and the afterglow emission of GRBs  (see Fig. \ref{fig:noM}). This relation strongly links the X-ray and $\gamma$-ray isotropic energy with the intrinsic peak energy, $E_{X,iso} \propto E_{\gamma,iso}^{1.00\pm0.06}/E_{p,i}^{0.60\pm0.10}$, unveiling an interesting connection between the early, more energetic  component of GRBs  ($E_{\gamma,iso}$ and $E_{p,i}$) and their late emission ($E_{X,iso}$), opportunely filtered for flaring activity. In addition, the intrinsic extra scatter is very small $\sigma_{E_X,iso} = 0.30 \pm 0.06$ ($1\sigma$). Starting from the results obtained by B12 and M13, and combining them with the well-known $E_{p,i}-E_{iso}$ relation \citep{Amati2002}, we present here a new correlation involving four GRB parameters, which we named the Combo-relation. This new relation is then used to measure  $\Omega_m$ and to explore the possible evolution of $w$ of EOS with the redshift by using as a ``candle'' the initial luminosity $L_0$ of the shallow phase of the afterglow, which we find to be strictly correlated with quantities directly inferred from observations.

The paper is organized as follows. In Sect. \ref{sec:2} we describe in detail how we obtain the formulation of the newly proposed GRB correlation, and present the sample of GRBs  used to test our results, the procedure for the fitting of the X-ray afterglow light curves, and finally, the existence of the correlation assuming a standard cosmological scenario. Then, in Sec. \ref{sec:3}, we discuss the use of the relation as cosmological parameter estimator, which involves  a calibration technique that does not require the use of the entire sample of SNe Ia, as was often done in previous similar works. In Sec. \ref{sec:5}, we test the use of GRBs estimating the main cosmological parameters, as well as the evolution of the dark energy EOS. Finally, we discuss the final results in the last section.

\section{The Combo-relation}\label{sec:2}

We present here a new GRB correlation, the Combo-relation, obtained after combining $E_{\gamma,iso}$--$E_{X,iso}$--$E_{p,i}$ (B12 and M13), the $E_{\gamma,iso}$--$E_{p,i}$ \citep{Amati2002} correlations, and the analytical formulation of the X-ray afterglow component given in \citet{RuffiniMuccino2014} (hereafter R14). 

The three-parameter scaling law reported in B12 and M13 can be generally written as
\begin{equation}\label{eq:no1}
E_{X,iso} \propto \frac{E_{\gamma,iso}^{\beta}}{E_{p,i}^{\delta}},
\end{equation}
where $E_{X,iso}$ is the isotropic energy of a GRB afterglow in the rest-frame ($0.3$--$30$) keV energy range obtained by integrating the light curve in luminosity over a specified time interval, $E_{\gamma,iso}$ is the isotropic energy of a GRB prompt emission, and $E_{p,i}$ is the intrinsic spectral peak energy of a GRB. Since $E_{X,iso}$ and $E_{\gamma,iso}$ are both cosmological-dependent quantities, we reformulate this correlation in order to involve only one cosmological-dependent quantity.

The right term of Eq. \ref{eq:no1} can be rewritten using the well-known formulation of the Amati relation, $E_{\gamma,iso}\propto E_{p,i}^{\eta}$, which provides
\begin{equation}\label{eq:no2}
E_{X,iso} = A\ E_{p,i}^{\gamma},
\end{equation}
where $\gamma=\eta\times\beta-\delta$ and $A$ is the normalization constant. Since the Amati relation is valid only for long bursts, in the following discussion we will exclude short GRBs with a rest-frame $T_{90}$ duration smaller than $2$ s, and short bursts with ``extended emission'' \citep{Norris2006}, or ``disguised short'' GRBs \citep{Bernardini2007,Caito2010}, which have hybrid characteristic between short and long bursts.

To rewrite Eq. \ref{eq:no2} for cosmological purposes, we have calculated the total isotropic X-ray energy in the rest-frame ($0.3$--$10$) keV energy range by integrating the X-ray luminosity $L(t^\prime_a)$, expressed as a function of the cosmological rest-frame arrival time $t^\prime_a$, over the time interval $(t^\prime_{a,1}, t^\prime_{a,2})$.
This luminosity $L(t^\prime_a)$ is obtained by considering four steps \citep[see e.g. Appendix \ref{app:a} and][]{Pisani2013}.

It is well known that the X-ray afterglow phenomenology can be described by the presence of an additional component emerging from the soft X-ray steep decay of the GRB prompt emission, and characterized by a first shallow emission, usually named the plateau, and a late power-law decay behaviour \citep{Nousek2006,Zhang2006,Willingale2007}. In addition, many GRB X-ray light curves are characterized by the presence of large, late--time flares, whose origin is very likely associated with late--time activity of the internal engine \citep{Margutti2010}. Since their luminosities are much lower than the prompt one, we exclude X--ray flares from our analysis via a light--curve fitting algorithm, which will be explained later in the text. We then make  the assumption that the $E_{X,iso}$ quantity refers only to the component whose X-ray luminosity $L(t^\prime_a)$ is given by the phenomenological function defined in R14
\begin{equation}
\label{eq:no3}
L(t_a^\prime) = L_0 \left( 1 + \frac{t^\prime_a}{\tau} \right)^{\alpha_X},
\end{equation}
where $L_0$, $\tau$, and $\alpha_X$ are, respectively, the luminosity at $t_a^\prime=0$, the characteristic timescale of the end of the shallow phase, and the late power-law decay index of a GRB afterglow.

Therefore, if we extend the integration time interval to $t^\prime_{a,1}\rightarrow0$ and $t^\prime_{a,2}\rightarrow+\infty$, the integral of the function $L(t^\prime_a)$ in Eq. \ref{eq:no3} gives
\begin{equation}
\label{eq:no4}
E_{X,iso} = \int_{0}^{+\infty} L(t^\prime_a)\ d t^\prime_a = L_0\ \frac{\tau}{|1 + \alpha_X |}\ , 
\end{equation}
with the requirement that $\alpha_X<-1$. 
This condition is necessary to exclude divergent values of $E_{X,iso}$ computed from Eq. (\ref{eq:no4}), for $t^\prime_{a,2}\rightarrow+\infty$.
It is worth  noting that light curves providing values $\alpha_X>-1$ could have a change in slope at very late times (beyond the XRT time coverage) and/or, in principle, could be polluted by a late flaring activity, resulting in a less steep late decay.

Considering Eqs. \ref{eq:no2} and \ref{eq:no4}, we can finally formulate the following relation between GRB observables
\begin{equation}\label{eq:no5}
L_0 = A E_{p,i}^{\gamma}\left(\frac{\tau}{|1 + \alpha_X |}\right)^{-1},
\end{equation}
which we name the Combo-relation.
At first sight, Eq. \ref{eq:no5} suggests the existence of a physical connection between specific physical properties of the afterglow and prompt emission in long GRBs. 
A two-dimensional fashion of the correlation in Eq. \ref{eq:no5} in logarithmic units can be written as
\begin{equation}
\label{eq:no7}
\log \left(\frac{L_0}{\textnormal{erg/s}}\right) = \log \left(\frac{A}{\textnormal{erg/s}}\right) + \gamma\left[ \log \left(\frac{E_{p,i}}{\textnormal{keV}}\right) - \frac{1}{\gamma}\log \left(\frac{\tau/\textnormal{s}}{|1+\alpha_X|}\right)\right]\ ,
\end{equation}
where the set of parameters in the square brackets has the meaning of a logarithmic independent coordinate, and in the following will be expressed as $\log \left[ X(\gamma,E_{p,i}, \tau, \alpha_X) \right]$.

We have tested the reliability of Eqs. \ref{eq:no5} and \ref{eq:no7} by building a sample of GRBs satisfying the following restrictions:

\begin{itemize}
\item a measured redshift $z$;
\item a determined prompt emission spectral peak energy $E_{p,i}$;
\item a complete monitoring of the GRB X-ray afterglow light curve from the early decay ($t^\prime_a \sim 100$ s, when present) until late emission ($t^\prime_a \sim 10^5$--$10^6$ s).
\end{itemize}

\begin{figure}
\centering
\includegraphics[width=\hsize,clip]{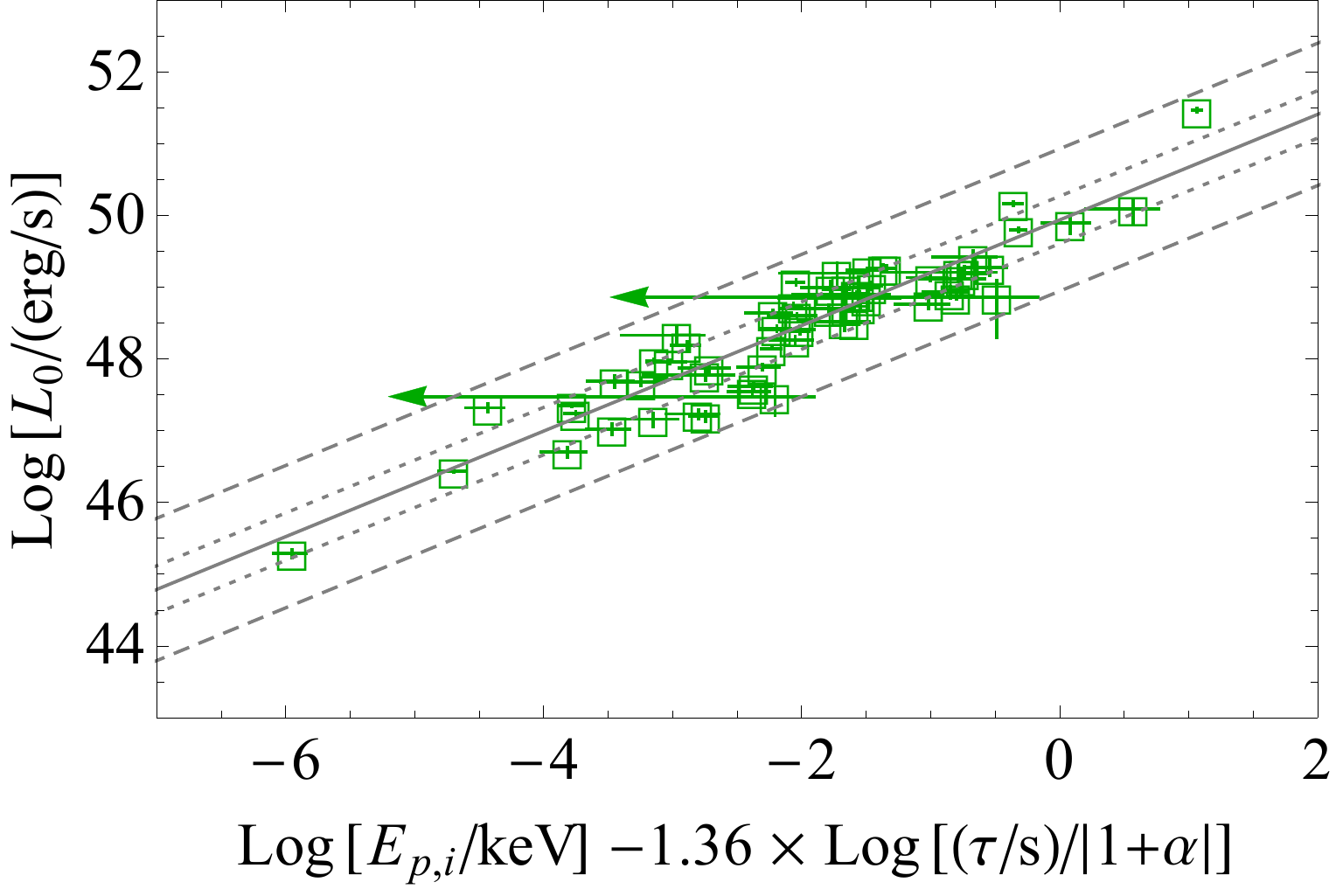}
\caption{Plot of the correlation considering the entire sample of $60$ GRBs. The green empty boxes are the data of each of the sources, derived as described in  Appendix \ref{app:b}, the solid black line is the best fit of the data, while the dotted grey lines and the dashed grey lines correspond, respectively, to the dispersion on the correlation at $1\sigma_{ex}$ and $3\sigma_{ex}$.}
\label{fig:no}
\end{figure}

We start the analysis by computing the rest-frame ($0.3$--$10$) keV energy band light curves (see Appendix \ref{app:a}). 
The fitting of the continuum part of the X-ray light curves was performed by using a semi-automated procedure based on the $\chi^2$ statistic, which eliminates the flaring part \citep[see e.g. Appendix \ref{app:b} and][and M13, for details]{Margutti2011,Zaninoni2013}. A total of 60 GRBs are found, whose distribution in the Combo-relation plane is shown in Fig. \ref{fig:no}, where the value of the luminosity $L_0$ for each GRB is calculated from the flat $\Lambda$CDM scenario. The corresponding best-fit parameters, as well as the extra-scatter term $\sigma_{ext}$, have been derived by following the procedure by \citet{Dago2005}, and are respectively $\log\left[A/(\textnormal{erg/s})\right] = 49.94\pm0.27$, $\gamma = 0.74\pm0.10$, and $\sigma_{ext} = $0.33$\pm$0.04. The Spearman's rank correlation coefficient is $\rho_S = 0.92$, while the p-value computed from the two-sided Student's $t$-distribution, is $p_{val} = 9.13\times10^{-22}$.

\section{Calibration of the Combo-relation}\label{sec:3}

\begin{figure}
\centering
\includegraphics[width=\hsize,clip]{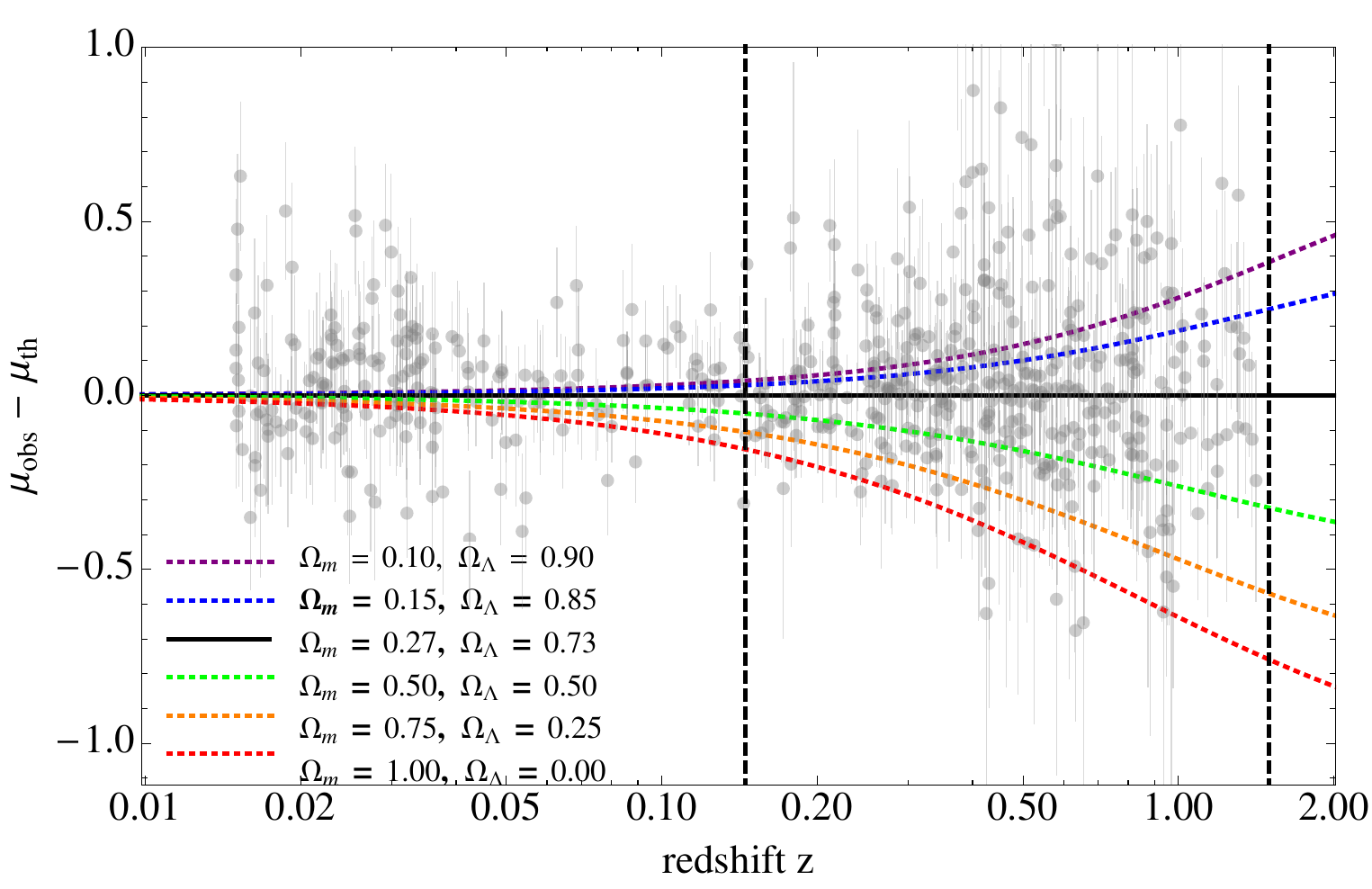}
\caption{The residual distance modulus $\mu_{obs} - \mu_{th}$ for different values of the density cosmological parameters $\Omega_m$ and $\Omega_{\Lambda}$ up to $z = 2.0$. We consider the best fit to be the standard $\Lambda$CDM model, where $\Omega_m = 0.27$, $\Omega_{\Lambda} = 0.73,$ and $H_0 = 71$ km/s/Mpc (black line). Union2 SNe Ia data residuals are shown in grey. The large spread (more than 1 magnitude) shown by  $\mu$ at $z = 1.5$ and at $z = 0.145$ (the two vertical dashed lines) where the scatter is almost 0.2 magnitudes is clearly evident.}
\label{fig:no3}
\end{figure}

The lack of very nearby ($z \sim 0.01$) GRBs prevents the possibility of calibrating GRB correlations, as  is usually done with SNe Ia. In recent years different methods have been proposed to avoid this ``circularity problem'' in calibrating GRB relations \citep[][and references therein]{Kodama2008,Demianski2012}. The common approach uses an interpolating function for the distribution of luminosity distances (distance moduli) of SNe Ia with redshift, and then extending it to GRBs at higher redshifts. In the following we introduce and describe an alternative method for calibrating the Combo-relation (see also \citealp{Ghirlanda2006}) which consists of two steps: 1)  we identify a small but sufficient subsample of GRBs that lie at the same redshift, and then infer the slope parameter $\gamma$ from a best-fit procedure; 2) once we determine $\gamma$, the luminosity parameter $A$ can be obtained from a direct comparison between the nearest, $z = 0.145$, GRBs in our sample and SNe Ia located at very similar redshifts. We note that this approach is different from previous ones in that we do not use the whole redshift range covered by SNe Ia, but we limit our calibration analysis to z = 0.145 where the effect of the cosmology on the distance modulus of the calibrating SNe Ia is small (see Fig. \ref{fig:no3}).

\subsection{The determination of the slope $\gamma$}

\begin{table*}
\centering
\caption{List of the five GRBs used for the determination of the slope parameter $\gamma$ of the correlation. In the Col. 1  is shown the GRB name, in Col. 2 the redshift $z$, in Col. 3  the intrinsic peak energy of the burst, in Col. 4  the initial flux of the afterglow additional component $F_0$ in the common rest-frame energy range (0.3 -- 10) keV, in Col. 5 the time parameter ($\tau$), in Col. 6 the late power-law decay index ($\alpha_X$).}
{
\begin{tabular}{lccccc}
\hline\hline
GRB      &  z         &  $E_{p,i}$ (keV)  &  $F_0$ (erg cm$^{-2}$s$^{-1}$)   &  $\tau$ (s)      & $\alpha_X$       \\
\hline
060729   &  $0.54$    &  $104\pm24$       &  $(2.32\pm0.05)\times10^{-11}$   &  $61462\pm2844$  &  $-1.71\pm0.03$  \\
081007   &  $0.5295$  &  $61\pm15$        &  $(4.48\pm0.64)\times10^{-11}$   &  $2178\pm457$    &  $-1.17\pm0.04$  \\
090424   &  $0.544$   &  $273\pm5$        &  $(4.20\pm0.26)\times10^{-9}$    &  $204\pm16$      &  $-1.20\pm0.01$  \\
090618   &  $0.54$    &  $257\pm41$       &  $(1.18\pm0.03)\times10^{-9}$    &  $1206\pm47$     &  $-1.46\pm0.01$  \\
100621A  &  $0.542$   &  $146\pm23$       &  $(1.75\pm0.22)\times10^{-10}$   &  $3613\pm746$    &  $-1.05\pm0.03$  \\
\hline
\end{tabular}
}
\label{tab:no2}
\end{table*}

The existence of a subsample with a sufficient number of GRBs lying at almost the same cosmological distance would, in principle, allow us to infer $\gamma$, overriding any possible cosmological dependence (assuming a homogeneous and isotropic universe). In our sample of $60$ GRBs there is a small subsample of $5$ GRBs located at a very similar redshift, see Table \ref{tab:no2}. The difference between the maximum redshift of this $5$--GRB sample $(z = 0.544)$ and the minimum one $(z = 0.5295)$ corresponds to a variation of $0.015$ in redshift and of $0.07$ in distance modulus $\mu$ in the case of the standard $\Lambda$CDM model. This very small difference is  sufficient for our purposes.

\begin{figure}
\centering
\includegraphics[width=\hsize,clip]{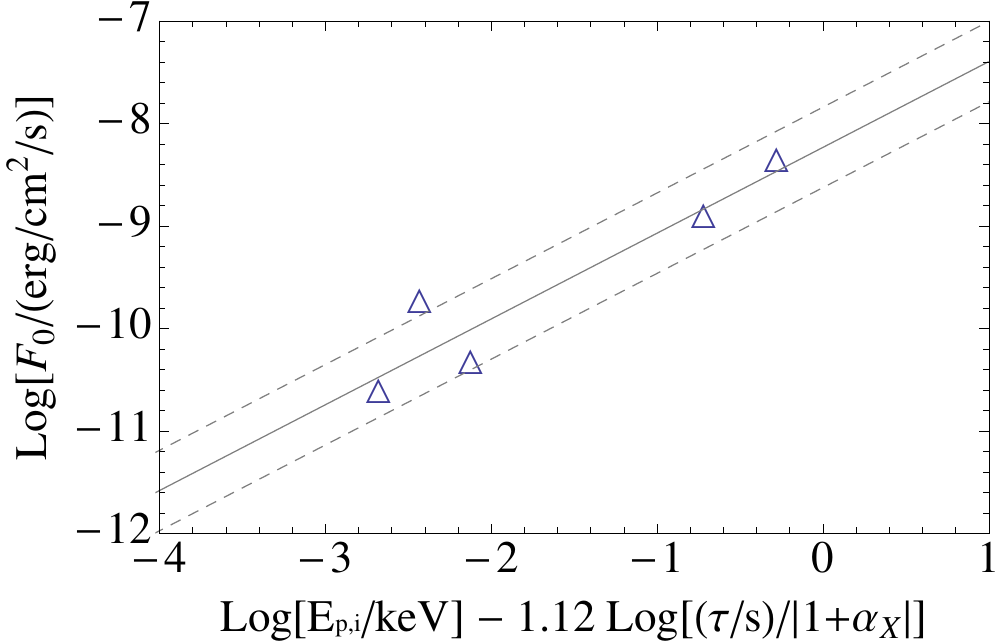}
\caption{Plot of the correlation found for the sample of five GRBs located at the same redshift. The blue triangles are the data of each of the five sources, derived from the procedure in Appendix \ref{app:b}. The solid line represents the best fit while the dashed line is the dispersion on the correlation at $1\sigma_{ex}$. }
\label{fig:no4}
\end{figure}

To avoid any possible cosmological contamination, we do not consider the luminosity $L_0$ as the dependent variable, but we consider instead the energy flux $F_0$, which is related to the luminosity through the expression $L_0 = 4 \pi d_l^2\ F_0$. This assumption does not influence the final result, since $d_l$ is almost the same for the $5$--GRB sample and, therefore, the $4\pi d_l^2$ term can be absorbed in the normalization constant. Consequently, we build the energy flux light curve for each GRB, and then, following the same procedure described in Sec. \ref{sec:2}, we perform a best-fit analysis of this subsample of five GRBs using the maximum likelihood technique.  From the best fit we
find a value of $\gamma=0.89\pm0.15$, with an extra scatter of $\sigma_{ext} = 0.40 \pm 0.04$, see Fig. \ref{fig:no4}. 

\subsection{Calibration of $A$ with SNe Ia}

Among the considered $60$ GRBs, the nearest one is GRB 130702A at $z = 0.145$. We use it to perform the calibration of the Combo-relation with SNe Ia located at the same distance. It is clear that at these redshifts ($z = 0.145$) the offset provided by SNe Ia is much smaller than the value inferred from SNe Ia at larger redshifts, see Fig. \ref{fig:no3}.
In this light we have selected Union2 SNe Ia \citep{Union2,Union2.1} with redshift between $z = 0.143$ and $z = 0.147$. We find five SNe Ia that satisfy this condition; their properties are shown in Table \ref{tab:no3}. We then compute the average value of their distance modulus and its uncertainty, $\langle\mu\rangle = 39.19 \pm 0.27$, which will be used for the final calibration. The parameter $A$  can be obtained inverting Eq. \ref{eq:no7}, and considering $L_0\ = 4 \pi dl^2\ F_0$:
\begin{align}
\nonumber
\log \left(\frac{A}{\textnormal{erg/s}}\right) =&\  2 \log \left(\frac{d_l}{cm}\right) + \log 4\pi + \log \left(\frac{F_0}{\textnormal{erg/cm$^2$/s}}\right) + \\
\label{eq:no8}
& - \gamma\log X(\gamma,E_{p.i},\tau,\alpha_X).
\end{align}
The generic SN distance modulus $\mu$ can be directly related to the luminosity distance $d_l$ by
\begin{equation}\label{eq:no9}
\mu = 25 + 5 \log \left(\frac{d_l}{\textnormal{Mpc}}\right) = -\ 97.45 + 5 \log \left(\frac{d_l}{\textnormal{cm}}\right) \, 
,\end{equation}
where the last equality takes into account the fact that $d_l$ is expressed in cm. Substituting this last expression for $ \log \left(d_l\right)$ in Eq. \ref{eq:no8} we obtain
\begin{equation}
\label{eq:no10}
\log \left(\frac{A}{\textnormal{erg/s}}\right) = \frac{2}{5 }\left(\langle\mu\rangle + 97.45\right) + \psi(\gamma,E_{p.i},\tau,\alpha_X,F_0)\ ,
\end{equation}
where the term $\psi(\gamma,E_{p.i},\tau,\alpha_X,F_0)$ comprises the last three terms on the right hand side of Eq. \ref{eq:no8}. Substituting the quantities for GRB 130702A in Eq. \ref{eq:no10}, and the value of $\langle\mu\rangle$ previously obtained, we infer a value for the parameter $ \log [A/(\textnormal{erg/s})] = 49.54 \pm 0.20$, where the uncertainty also takes   the $\sigma_{ext}$ value found above into account.

\begin{table}
\centering
\caption{List of the $5$ SNe Ia selected from the Union2 sample \citep{Union2,Union2.1} and used for the calibration of the GRB correlation. In the first column it is shown the name of the SN, in the second column the redshift and in the third the distance modulus obtained from the light curve fitting with the SALT2 SED method \citep{SALT2}.}
{
\begin{tabular}{lcc}
\hline\hline
SN  &  z  &  $\mu$          \\
\hline
1999bm  &  0.1441  &  $38.836 \pm 0.157$   \\ 
10106  &  0.14629  &  $39.559 \pm 0.120$   \\ 
2005ln  &  0.14567  &  $39.057 \pm 0.126$   \\ 
2005gx  &  0.144621  &  $39.291 \pm 0.113$   \\ 
2005ld  &  0.1437  &  $39.186 \pm 0.116$   \\ 
\hline
\end{tabular}
}
\label{tab:no3}
\end{table}

\section{Cosmology with the Combo-relation}\label{sec:5}

\subsection{Building the GRB Hubble diagram}\label{sec:5.1}

We now discuss the possible use of the proposed GRB Combo-relation to measure the cosmological constant and the mass density, as well as their evolution with  redshift $z$.  The possibility of estimating the luminosity distance $d_l$ from the GRB observable quantities allows to us define a distance modulus for GRBs, and then its uncertainty, as 
\begin{equation}
\label{eq:no11}
\mu_{GRB} = - 97.45 + \frac{5}{2} \left[\mathcal{A} - \psi(\gamma,E_{p,i},\tau,\alpha_X,F_0) \right],
\end{equation}
where $\mathcal{A}=\log [A/(\textnormal{erg/s})]$. The quantity $\mu_{GRB}$ can be directly compared with the theoretical cosmological expected value $\mu_{th}$, which depends on the density parameters $\Omega_{m}$ and  $\Omega_{\Lambda}$, the curvature term $\Omega_k = 1-\Omega_m-\Omega_\Lambda$, and the Hubble constant $H_0$
\begin{equation}
\label{eq:no12}
\mu_{th} = 25 + 5 \log d_l(z,\Omega_m,\Omega_{\Lambda},H_0)\ .
\end{equation}
The luminosity distance $d_l$ is defined as
\begin{equation}
\label{eq:no13}
d_l = \frac{c}{H_0}(1+z)
\left\{ \begin{array}{lll} 
\frac{1}{\sqrt{| \Omega_k |}} \sinh \left[\int_0^z \frac{\sqrt{| \Omega_k |} dz^\prime}{\sqrt{E(z^\prime,\Omega_m,\Omega_{\Lambda}, w(z))}}\right] \hspace{-0.1cm}&,\  \Omega_k > 0 \vspace{0.2cm}\\
\int_0^z \frac{dz^\prime}{\sqrt{E(z^\prime,\Omega_m,\Omega_{\Lambda}, w(z))}} \hspace{-0.1cm}&,\ \Omega_k = 0 \vspace{0.2cm}\\ 
\frac{1}{\sqrt{| \Omega_k |}} \sin \left[ \int_0^z \frac{\sqrt{| \Omega_k |}dz^\prime}{\sqrt{E(z^\prime,\Omega_m,\Omega_{\Lambda}, w(z))}} \right] \hspace{-0.1cm}&,\  \Omega_k < 0 
\end{array} \right.
\end{equation}
with $E(z,\Omega_m,\Omega_{\Lambda}, w(z))=\Omega_m (1+z)^3 + \Omega_{\Lambda} (1+z)^{3(1+w(z))} + \Omega_k (1+z)^2$ \citep[see e.g.][]{Goobar1995}. 
In the following we fix the Hubble constant at the recent value inferred from low-redshift SNe Ia, corrected for star formation bias, and calibrated with the LMC distance \citep{Rigault2014}: $H_0 = 70.6 \pm 2.6$.

\begin{figure}
\centering
\includegraphics[width=\hsize,clip]{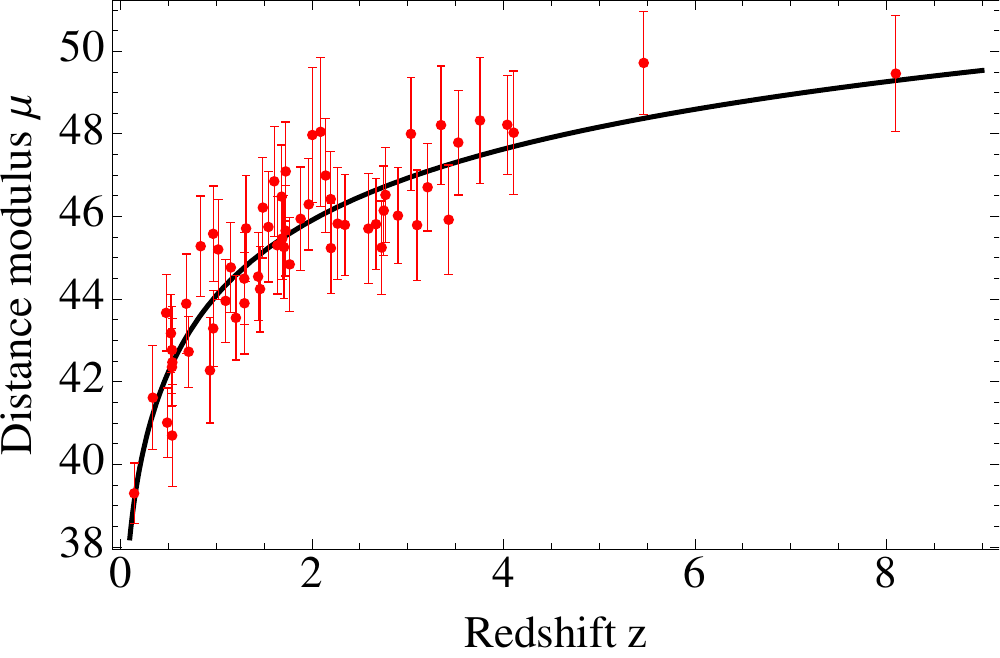}
\caption{The Combo-GRB Hubble diagram. The black line represents the best fit for the function $\mu(z)$ as obtained by using only GRBs and for the case of the $\Lambda$CDM scenario. }
\label{fig:no7}
\end{figure}

The corresponding uncertainties on $\mu_{GRB}$ were computed considering an observed term, $\sigma \mu_{obs}$, which takes into account each uncertainty on the observed quantities of the Combo-relation, e.g. $F_0, \tau, \alpha$, and $E_{p,i}$, and a ``statistical'' term, $\sigma \mu_{rel}$, which takes into account  the uncertainties on the parameters of the Combo-relation, $A$ and $\gamma$, and the weight of the extra scatter value $\sigma_{ext}$. The final uncertainty on each single GRB distance modulus
\begin{equation}
\label{eq:no15}
\sigma \mu_{GRB} = \sqrt{(\sigma \mu_{obs})^2 + (\sigma \mu_{rel})^2} 
\end{equation}
allows us to build the Combo-GRB Hubble diagram \citep[see also][]{Izzo2009}, which is shown in Fig. \ref{fig:no7}. It is possible to quantify the reliability of any cosmological model with our sample of 60 GRBs, which represents a unique dataset from low redshift ($z = 0.145$) to very large distances ($z = 8.23$). To reach our goal, we make the fundamental assumption that our GRB sample is normally distributed around the best-fit cosmology, which we are going to estimate. With this hypothesis, we consider as test-statistic the chi-square test, which is defined as 
\begin{equation}
\label{eq:no16}
\chi^2 = \sum_{i=1}^{60} \frac{\left[\mu_{GRB,i}(z, \mathcal{A}, \psi) - \mu_{th}(z, \Omega_m, \Omega_{\Lambda}, w(z))\right]^2}{\sigma \mu_{GRB,i}^2},
\end{equation}
where $\mu_{GRB}(z, \mathcal{A}, \psi)$, $\mu_{th}(z, \Omega_m, \Omega_{\Lambda}, w(z)),$ and $\sigma \mu_{GRB}$ are respectively defined in Eqs. \ref{eq:no11}, \ref{eq:no12}, and \ref{eq:no15}. The cosmology is included in the quantity $\mu_{th}(z, \Omega_m, \Omega_{\Lambda}, H_0)$, which we allow to vary. To determine the best configuration of parameters that most closely fits  the distribution of GRBs in the Hubble diagram we maximize the log-likelihood function, $-2\ln(e^{\chi^2})$, which is equivalent to the minimization of the function defined in Eq. \ref{eq:no16}.

\subsection{Fit results}

\subsubsection{$\Lambda$CDM case}

\begin{figure}
\centering
\includegraphics[width=\hsize,clip]{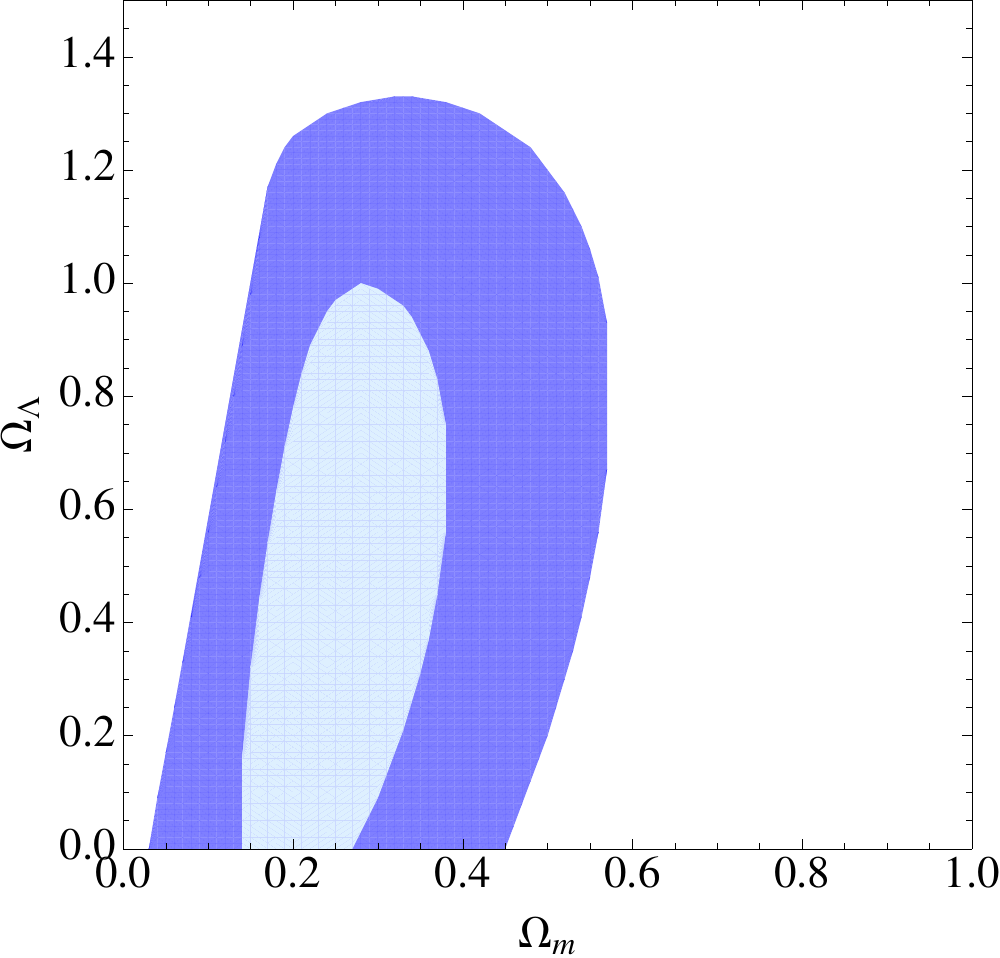}
\caption{ 1-$\sigma$ $(\Delta \chi^2 = 2.3)$ confidence region in the ($\Omega_{\rm{m}}$, $\Omega_{\Lambda}$) plane \textit{(left panel)} and in the ($\Omega_{\rm{m}}$, $w_0$) plane \textit{(right panel)} for the Combo-GRB sample \textit{(dark blue)}, and for the total ($60$ observed + $300$ MC simulated GRBs, \textit{(light blue)}.}
\label{fig:no8_0}
\end{figure}

In the $\Lambda$CDMmodel, the energy function $E(z,\Omega_m,\Omega_{\Lambda}, w(z))$ is characterized by an EOS for the dark energy term fixed to $w = -1$. Since we have that $\Omega_m + \Omega_{\Lambda} + \Omega_k =1$, we vary the matter and cosmological constant density parameters, also obtaining in this way  an estimate of the curvature term. We obtain that GRBs alone provide $\Omega_m = 0.29^{+0.23}_{-0.15}$, see also Fig. \ref{fig:no8_0}.

\subsubsection{Variable $w_0$ case}

In a flat Universe ($\Omega = \Omega_m + \Omega_{\Lambda} = 1$) with a constant value of $w_0$ different from the standard value $w = -1$, we can provide useful constraints for alternative dark energy theories. In this case, we only vary  the matter density and the dark energy equations of state, obtaining an estimate of the density matter of $\Omega_m = 0.40^{+0.22}_{-0.16}$ and of a dark energy EOS parameter $w_0 = -1.52^{+0.94}_{-0.93}$.

\begin{figure}
\centering
\includegraphics[width=\hsize,clip]{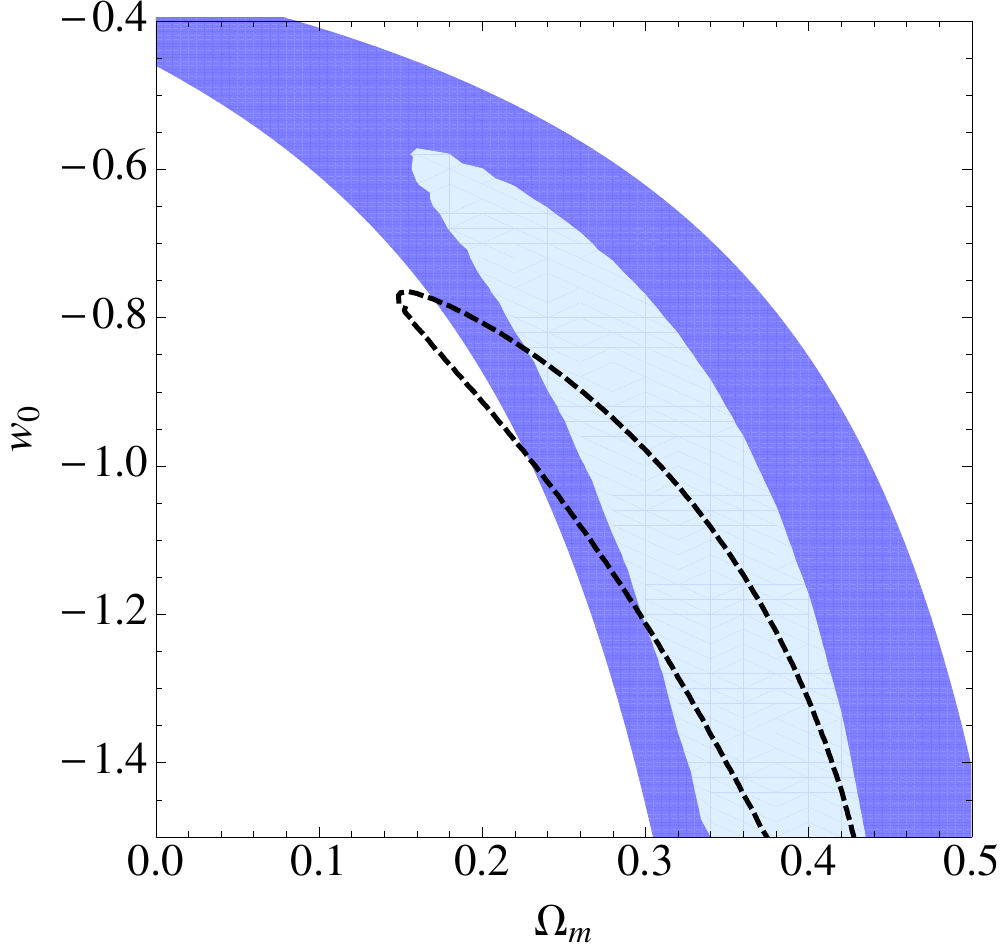}
\caption{ 1-$\sigma$ $(\Delta \chi^2 = 2.3)$ confidence region in the ($\Omega_{\rm{m}}$, $w_0$) plane \textit{(right panel)} for the Combo-GRB sample \textit{(dark blue)}, and for the total of $60$ observed + $300$ MC simulated GRBs \textit{(light blue)}. The black dashed line represents the 1-$\sigma$ confidence region obtained using the recent Union 2.1 SNe Ia sample \citep{Union2.1}.}
\label{fig:no8_1}
\end{figure}

\subsubsection{Evolution of $w(z)$} 

An interesting case--study consists of a time-evolving dark energy EOS in a flat cosmology, since the evolution of $w(z)$ can be directly studied with GRBs at larger redshifts. We consider an analytical formulation for the evolution of $w$ with the redshift, which was proposed by \citet{Chevallier2001} and \citet{Linder2003} (CPL), and where the $w(z)$ can be parameterized by

\begin{equation}\label{eq:wz}
w(z) = w_0 + w_a \frac{z}{1+z}.
\end{equation}

The CPL parameterization implies that for large $z$ the $w(z)$ term tends to the asymptotic value $w_0 + w_a$. Using a sample extending at large redshifts, e.g.  GRBs, will allow a better estimate of these parameters since the effects of a varying $w(z)$ on the distance modulus are more evident at  redshift $z \geq 1$. The GRB data provide a best-fit result $(w_0 = -1.43^{+0.78}_{-0.66}, w_a = 1.87^{+1.38}_{-2.57})$. 

\subsubsection{ Perspective: a Monte Carlo simulated sample of GRBs}

The current sample of GRBs that satisfies the Combo-relation is quite limited ($60$ bursts), when compared to the sample in \citet{Amati2013} ($\sim200$ bursts) or the SNe sample in the Union 2.1 release \citep{Union2.1}. A more numerous sample can help to understand whether the Combo-relation can provide better constraints on the cosmological parameters. 
To this aim, following the prescription of \citet{Li2007}, we used Monte Carlo (MC) simulations to generate a sample of $300$ synthetic GRBs satisfying the Combo-relation. This value comes from the expected number of GRBs detected in five years of operations of current (Swift) and future (SVOM \citep{SVOM}, and LOFT \citep{LOFT2012}) missions dedicated to observing GRBs.

First, we fitted the log-normal distributions of the $60$ observed $z$, $E_{p,i}$, $\tau$, and $|\alpha+1|$,
\begin{equation}
\label{eq:lognormal}
f(\log \xi) = \frac{1}{\sqrt{2\pi}\sigma_\xi}\exp\left[-\frac{\left(\log \xi-\mu_\xi\right)^2}{2\sigma_\xi^2}\right]\ ,
\end{equation}
where $\xi=z$, $E_{p,i}$, $\tau$, and $|\alpha+1|$, and we found the following mean values and dispersions: $\mu_z\pm\sigma_z=0.26\pm0.27$, $\mu_{E_{p,i}}\pm\sigma_{E_{p,i}}=2.54\pm0.40$, $\mu_\tau\pm\sigma_\tau=2.85\pm0.65$, and $\mu_{|\alpha+1|}\pm\sigma_{|\alpha+1|}=-0.36\pm0.31$.
Then, from these distributions we computed $\log X(\gamma,E_{p.i},\tau,\alpha)$ and we generated the initial luminosity $\log L_0=\gamma \log X+A$ from the frequency distribution
\begin{equation}
\label{eq:lognormal}
f(\log L_0) = \frac{1}{\sqrt{2\pi}\sigma_{ext}}\exp\left[-\frac{\left(\log L_0 - \gamma \log X -A \right)^2}{2\sigma_{ext}^2}\right]\ ,
\end{equation}
assuming that the Combo-relation is independent of the redshift and considering its extra-scatter $\sigma_{ext}$ . The values of $\gamma$, $A$, and $\sigma_{ext}$ are reported in Section 3.
Finally, to complete the set of parameters necessary to compute the distance modulus of the simulated sample of GRBs from each pair ($\log L_0$, $z$), we generated the corresponding $\log F_0$ ($\mu_{F_0}\pm\sigma_{F_0}=-9.87\pm0.85$). In the following, the attached errors on the MC simulated Combo-relation parameters will be taken as  $30\%$ of their corresponding values, which reflects the uncertainty of the ``real'' GRB sample.

We have verified whether the constraints on ($\Omega_{\rm{m}}$, $\Omega_{\Lambda}$) will improve by using a larger sample of 360 GRBs (the real sample of $60$ GRBs observed and a MC-simulated sample of $300$ GRBs, described above). The improvement on the constraints on $\Omega_{\rm{m}}$ and $\Omega_{\Lambda}$ is clear: the uncertainties on the density parameters improve considerably ($\Omega_m = 0.27^{+0.09}_{-0.05}$ for the $\Lambda$CDM case, $w_0 = -1.16^{+0.32}_{-0.38}$ for the variable $w_0$ case), as  is also clear in the contour plots shown in Figs. \ref{fig:no8_0} and \ref{fig:no8_1}.

\subsection{GRBs compared with SNe, CMBs, BAOs}\label{sec:4.4}

In order to compare the Combo-GRB sample results, we also consider  the following datasets:

\begin{itemize}
\item[-] the measurements of the baryon acoustic peaks $A_{obs} = (0.474 \pm 0.034, 0.442 \pm 0.020, 0.424 \pm 0.021)$ at the corresponding redshifts $z_{BAO} = (0.44, 0.6, 0.73)$ in the galaxy correlation function as obtained by the WiggleZ dark energy Survey \citep{WiggleZ}. The BAO peak is defined as  \citep{Eisenstein2005}
\begin{equation}
\label{eq:no17}
A_{BAO} = \left[ \frac{cz_{BAO}}{H_0}\frac{r(z_{BAO},\Omega_m,\Omega_\Lambda,w(z))^2}{E(z,\Omega_m,\Omega_\Lambda,w(z))}\right]^{1/3} \frac{\sqrt{\Omega_m H_0^2}}{c z_{BAO}}
,\end{equation}
where $r(z,\Omega_m, \Omega_{\Lambda}, w(z))$ is the comoving distance. The best-fit cosmological model is determined by the minimization of the corresponding chi-squared quantity
\begin{equation}
\label{eq:no18}
\chi^2_{BAO} = \sum_{i=1}^3 (A_{BAO} - A_{(obs,i)})^T C^{-1} (A_{BAO} - A_{(obs,i)}),
\end{equation}
where $C^{-1}$ is the inverse covariance matrix of the measurements of the WiggleZ survey \citep{WiggleZ}.
\item[-] the measurement of the shift parameter $R_{obs} = 1.7407 \pm 0.0094$ as obtained from the Planck first data release \citep{Planckfirstrelease}. The $R$ quantity is the least cosmological model-dependent parameter (particularly from $H_0$) that can be extracted from the analysis of the CMB \citep{WangMukherjee2006} and is defined as
\begin{equation}
\label{eq:no19}
R = \sqrt{\Omega_m} \int_0^{z_{rec}} \frac{d z'}{E(z',\Omega_m, \Omega_{\Lambda}, w(z))},
\end{equation}
where $z_{rec}$ is the redshift of the recombination. The best-fit cosmological model is determined by the minimization of the corresponding chi-squared term 
\begin{equation}
\label{eq:no20}
\chi^2_{CMB} = \frac{(R_{obs} - R)^2}{\sigma R_{obs}^2}.
\end{equation}
\end{itemize}

\begin{figure}
\centering
\includegraphics[width=\hsize,clip]{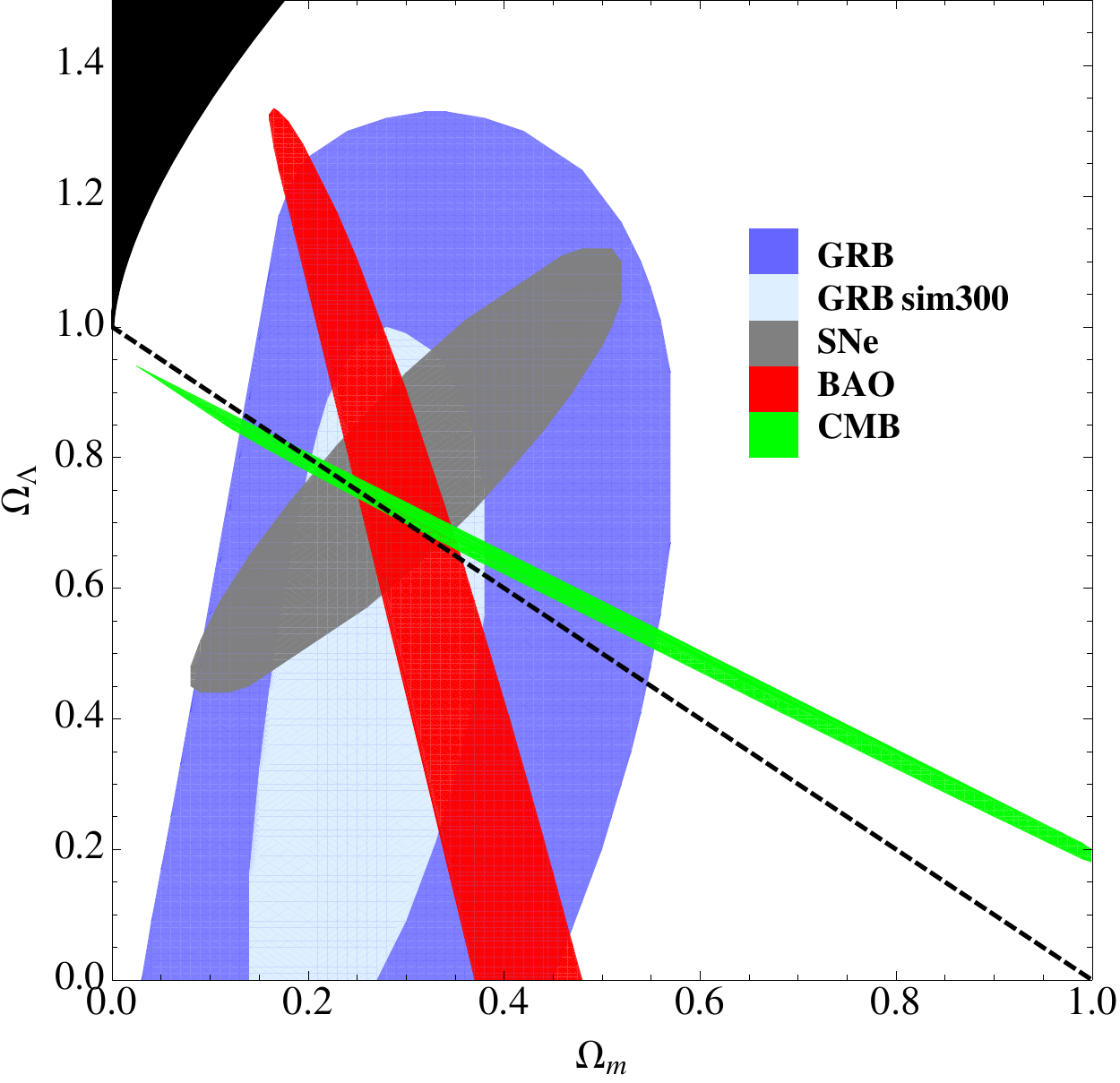}
\caption{1-$\sigma$ $(\Delta \chi^2 = 2.3)$ confidence region in the ($\Omega_{\rm{m}}$, $\Omega_{\Lambda}$) plane for the observed GRB sample (\textit{blue}), with the inclusion of the MC--simulated 300 GRBs sample (\textit{cyan}), and with the samples of SNe (\textit{grey}), BAOs (\textit{red}), and CMBs (\textit{green}). The dashed line represents the condition of the Flat Universe $\Omega_{\rm{m}}+\Omega_{\Lambda}=1$.}
\label{fig:no8}
\end{figure}

We use a grid-search technique to vary the values of the cosmological which are used to solve numerically the chi-squared equations defined above. Every dataset provides the respective distribution of the cosmological parameters, so when combining datasets we simply determine the best-fit model as the sum of the single $\chi^2_i$ values for any pair of parameters and for the combined values, 
\begin{equation}
\label{eq:no21}
\chi^2_{tot} = \chi^2_{GRB} + \chi^2_{SNe} + \chi^2_{BAO} + \chi^2_{CMB},
\end{equation}
 and make the final plots using the \textit{Mathematica}\footnote{http://www.wolfram.com/mathematica/} software suite. 
Cosmological parameter uncertainties were estimated from single and combined $\chi^2$ statistics for each dataset. In Fig. \ref{fig:no8} we show the parameter spaces with the contour at the 1-$\sigma$ ($\Delta \chi^2 = 2.3$) confidence limit for each  considered dataset, and for the $\Lambda$CDM case.  We note that in all the datasets and fits, the Hubble constant is fixed to the value of $H_0 = 70.6 \pm 2.6$ \citep{Rigault2014}. From the combination of all the considered datasets, for the $\Lambda$CDM case, we obtain a matter density value of $\Omega_m = 0.28^{+0.01}_{-0.01}$, which shows no strain with the Combo-GRB results and the simulated one, see Fig. \ref{fig:no8}.

\section{Discussions and conclusions}\label{sec:7}

In this paper we have presented the ``Combo'' relation, a new tool for GRB cosmology. This relationship provides a very close link between prompt and afterglow parameters and it is characterized by a small intrinsic scatter, which makes this correlation very suitable for cosmological purposes. We recognize the fundamental role of the \textit{Swift}-XRT \citep{Gehrels2004,Burrows2005} which, thanks to its ability to slew very rapidly toward the location of a GRB event, provides real-time and detailed data of GRB afterglow light curves, whose evolution is at the base of the proposed Combo-relation. From our analysis the following results emerge:

\begin{itemize}
\item the proposed two--step calibration of the Combo--GRB relation greatly minimizes the dependence on SNe Ia;
\item GRBs data alone provide for the $\Lambda$CDM case $\Omega_m = 0.29^{+0.23}_{-0.15}$, see Fig. \ref{fig:no8};
\item a recent  paper   \citep{Milne2015} highlights the existence of an observational bias (a systematic difference in the velocity of SNe Ia ejecta, which is reflected in their curves), potentially affecting the measurements of cosmological parameters obtained with SNe Ia. On the basis of our results we conclude that given the current accuracy of GRB measurements we cannot exclude, within the errors,  that an effect like this is at play; however, this effect should not change the conclusions derived from SNe-Ia observations;
\item the launch of advanced and more sensitive detectors, such as the incoming SVOM \citep{SVOM} and the proposed LOFT \citep{LOFT2012} missions (and the expected Swift operations in the near future), will dramatically increase the number of GRBs in the dataset. In five years of operation of the SVOM mission alone, we expect to reach a ``good enough'' sample of 300 GRBs. With a Monte Carlo simulated sample of 300 GRBs, we will  significantly improve the accuracy of  $\Omega_m$ measurement, up to $\Omega_m = 0.27^{+0.09}_{-0.05}$, which is comparable with type Ia SNe \citep{Union2.1};
\item by using the CPL analytical parameterization, adopted to study the evolution of the dark energy EOS (see Eq. \ref{eq:wz}), we find $\Omega_m = 0.40^{+0.22}_{-0.16}$, $w_0 = -1.43^{+0.78}_{-0.66}$, and $w_a = 1.87^{+1.38}_{-2.57}$; 
\item the analysis of a combined (SNe+BAO+CMB) dataset confirms that the increasing size of the GRB sample will improve the accuracy of the measurement of the $\Omega_m$ parameter and in particular of the evolution of $w$ up to $z \sim$ 10.
\item the analytical expression of the Combo-relation provides an explicit close link, characterized by a small intrinsic scatter, between the prompt and the afterglow GRB emissions. This  points out the existence of a physical connection between the prompt and the afterglow emissions, which represents a new challenge for GRB models.
\end{itemize}

\begin{acknowledgements}

We thank the referee for her/his constructive comments which have improved the paper. EZ acknowledges the support by the International Cooperation Program CAPES-ICRANet financed by CAPES - Brazilian Federal Agency for Support and Evaluation of Graduate Education within the Ministry of Education of Brazil. This work made use of data supplied by the UK \emph{Swift} Science Data Centre at the University of Leicester. 
\end{acknowledgements}

\bibliographystyle{aa}

\appendix

\section{Computation of the rest-frame $0.3$--$10$ keV luminosity $L(t^\prime_a)$}\label{app:a}

The rest-frame $0.3$--$10$ keV luminosity $L(t^\prime_a)$ was obtained by considering four steps. 
\begin{itemize}
\item[(1)] We  obtained the \textit{Swift}-XRT flux light curves in the observed $0.3$--$10$ keV energy band\footnote{http://www.swift.ac.uk/burst\_analyser/}.
\item[(2)] We  transformed the observed flux $f_{obs}$ from the observed energy band $0.3$--$10$ keV to the rest-frame energy band $0.3$--$10$ keV by assuming an absorbed power-law function as the best fit for the spectral energy distribution of the XRT data, $N(E)\sim E^{-\gamma}$, with Galactic and intrinsic column densities obtained from the \ion{H}{I} radio map \citep{Kalberla2005} and from the best fit of the total afterglow spectrum, respectively. By using the photon indexes inferred for each time interval, the rest-frame flux light curve $f_{rf}$ is given by 
\begin{equation}
f_{rf}=f_{obs}\frac{\int^{10/(1+z)}_{0.3/(1+z)} N(E)E dE}{\int^{10}_{0.3} N(E) E dE}=f_{obs}(1+z)^{\gamma-2}.
\end{equation}
\item[(3)] We  transformed the observed time $t_a$ into the rest-frame time by correcting for $z$
\begin{equation}
t_{a}^\prime = t_a/(1+z).
\end{equation}
\item[(4)] We  defined the isotropic luminosity as
\begin{equation}
L=4 \pi d_l^2 f_{rf}.
\end{equation}
\end{itemize}

\section{Determination of the sample and verification of the correlation}\label{app:b}

To obtain the parameters involved in the Eq. \ref{eq:no5}, we needed to select an adequate sample of GRBs, to fit their X-ray light curves, and to collect or to calculate their $E_{p,i}$ values. The selection criteria have  already been delineated in Sec. \ref{sec:2}.

The entire procedure works in the rest-frame $0.3$--$10$ keV energy range for all GRBs. For the X-ray light curve fitting technique we developed a semi-automated code performing all the needed operations. The code is based on the IDL\footnote{Interactive Data Language, http://www.exelisvis.com/language/en-US/ProductsServices/IDL.aspx} language, and the fitting routine used is \textit{MPFIT}\footnote{http://purl.com/net/mpfit} \citep{Markwardt2009}, which is based on the non-linear least squares fitting. First, the procedure fits the complete light curve, then it eliminates at every iteration the data point with the largest positive residual, until  it obtains a fit with a p-value greater than $0.3$. To fit the light curves considered in luminosity units (erg/s), we use the composite function (R14):
\begin{enumerate}
\item a power law for the early steep decay
\begin{equation}\label{eq:no6}
L(t)=L_p \left(\frac{t^\prime_a}{100}\right)^{-\alpha_p},
\end{equation}
with $L_p$ the normalization factor and $\alpha_p$ the slope;
\item Eq. \ref{eq:no3} for the afterglow additional component.
\end{enumerate}

\begin{figure}
\centering
\includegraphics[width=\hsize,clip]{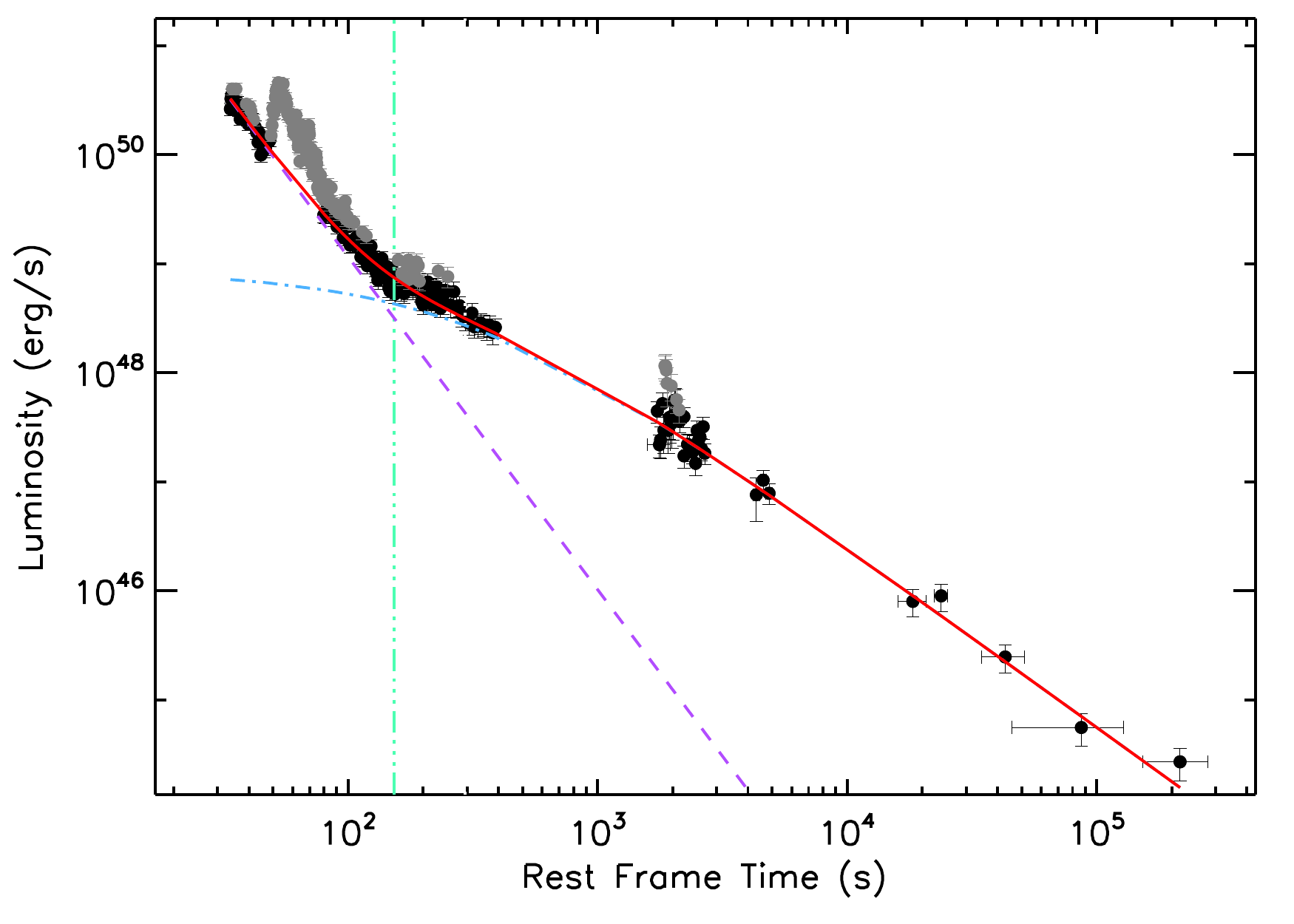}
\caption{Example of the combined fitting procedure (solid red line) as described in Eqs. \ref{eq:no3} and \ref{eq:no6}, filtered by the flares. The early steep decay fitted by using the power-law function in Eq. \ref{eq:no6} is indicated by the purple dashed line, while the afterglow additional component is fitted by the phenomenological function in Eq. \ref{eq:no3} (see also R14), and described by the dot-dashed cyan curve. In this specific case,  the luminosity light curve of GRB 060418 is shown in which the black dots with the error bars are the flare-free data, the grey dots are the excluded data recognized as due to the flares. The vertical green dotted line indicates the characteristic timescale of the parameter $\tau$.}
\label{fig:noex}
\end{figure}

An application of this joint fitting procedure is shown in Fig. \ref{fig:noex} for the case of GRB 060418. 
After the fitting procedure, we select only the GRBs with $\alpha_X<-1$, a condition necessary for the convergence of the integral in  Eq. \ref{eq:no4}.
The final total sample, summarized in Table \ref{tab:no}, consists of $60$ GRBs.

\begin{table*}
\centering
\tiny{
\caption{Long bursts with $\alpha_X<-1$ analysed in this work (first column) and their main parameters: GRB name (first column), redshift $z$ (second column), intrinsic peak energy $E_{p,i}$ (third column), flux in the $0.3$--$10$ keV rest-frame energy band $F_0$ (fourth column), the time parameter $\tau$ (fifth column), and the late power-law decay index $\alpha_X$ (sixth column). All errors are at the 1-$\sigma$ confidence level.}
\label{tab:no}
\begin{tabular}{lcccccccc}
\hline\hline
GRB      &  z        &  $E_{p,i}$ (keV)  &  $F_0$ (erg cm$^{-2}$s$^{-1}$)  &  $\tau$ (s)      &  $\alpha_X$ \\
\hline\hline
050401   &  $2.9$     &  $467\pm110$     &  $(2.35\pm0.13)\times10^{-10}$  &  $733\pm102$     &  $-1.64\pm0.07$   \\
050922C  &  $2.198$   &  $415\pm111$     &  $(3.65\pm0.39)\times10^{-10}$  &  $234\pm35$      &  $-1.51\pm0.04$   \\
051109A  &  $2.346$   &  $539\pm200$     &  $(2.24\pm0.29)\times10^{-10}$  &      $429\pm62$      &  $-1.26\pm0.02$   \\
060115   &      $3.53$    &  $285\pm34$      &  $(5.07\pm0.84)\times10^{-12}$  &  $12553\pm6079$  &  $-2.87\pm0.72$    \\
060418   &  $1.489$   &  $572\pm143$     &  $(6.00\pm1.10)\times10^{-10}$        &  $294\pm62$      &  $-1.65\pm0.06$   \\
060526   &  $3.21$    &  $105\pm21$      &  $(5.05\pm0.49)\times10^{-12}$  &  $25847\pm9019$  &  $-4.44\pm0.97$  \\
060707   &  $3.43$    &  $279\pm28$      &  $(1.94\pm0.53)\times10^{-11}$  &  $1067\pm485$    &  $-1.11\pm0.06$   \\
060729   &  $0.54$    &  $104\pm24$      &  $(2.28\pm0.05)\times10^{-11}$  &  $61462\pm2844$  &  $-1.71\pm0.03$   \\
060814   &  $0.84$    &  $473\pm155$       &  $(5.04\pm0.75)\times10^{-11}$  &      $3183\pm682$    &  $-1.30\pm0.05$   \\
060908   &  $1.8836$    &  $514\pm102$     &  $(1.02\pm0.25)\times10^{-9}$   &  $207\pm72$      &   $-1.67\pm0.12$   \\
060927   &  $5.467$   &  $475\pm47$      &  $(1.74\pm0.23)\times10^{-11}$  &  $230\pm73$      &  $-1.44\pm0.16$   \\
061121   &  $1.314$   &  $1289\pm153$    &  $(3.08\pm0.10)\times10^{-10}$  &  $1544\pm83$     &  $-1.54\pm0.02$    \\
071020   &  $2.145$       &  $1013\pm160$    &  $(3.50\pm1.30)\times10^{-9}$   &  $9.9\pm4.0$     &  $-1.16\pm0.02$   \\
080319B  &  $0.937$   &  $1261\pm65$     &  $(6.52\pm0.37)\times10^{-7}$   &     $22.11\pm0.97$  &  $-1.70\pm0.01$  \\
080413B  &  $1.1$     &  $150\pm30$      &  $(4.01\pm0.35)\times10^{-10}$  &  $184\pm24$      &  $-1.11\pm0.02$   \\
080605   &  $1.6398$  &  $650\pm55$      &  $(3.48\pm0.16)\times10^{-9}$   &  $83.7\pm6.0$    &  $-1.42\pm0.02$   \\
080607   &  $3.036$   &  $1691\pm226$    &  $(2.31\pm0.54)\times10^{-10}$  &  $387\pm91$      &  $-1.65\pm0.06$   \\
080721   &  $2.591$   &  $1741\pm227$    &  $(2.61\pm0.13)\times10^{-9}$   &  $431\pm100$     &  $-1.61\pm0.02$  \\
080810   &  $3.35$    &  $1470\pm180$    &  $(1.27\pm0.50)\times10^{-10}$  &  $647\pm239$     &  $-1.83\pm0.10$    \\
080916A  &  $0.689$   &  $184\pm18$      &  $(5.05\pm0.84)\times10^{-11}$  &  $967\pm251$     &  $-1.06\pm0.03$ \\
080928   &  $1.692$       &  $95\pm23$       &  $(3.83\pm0.56)\times10^{-11}$  &  $5976\pm1499$   &  $-3.11\pm0.33$   \\
081007   &  $0.5295$  &  $61\pm15$       &  $(4.61\pm0.66)\times10^{-11}$  &  $2178\pm457$    &  $-1.17\pm0.04$   \\
081008   &  $1.9685$  &  $261\pm52$      &  $(6.46\pm0.79)\times10^{-11}$  &  $1666\pm368$    &  $-1.86\pm0.13$   \\
081222   &  $2.77$        &  $505\pm34$      &  $(1.58\pm0.18)\times10^{-10}$  &  $699\pm93$      &  $-1.61\pm0.04$  \\
090102   &  $1.547$   &  $1149\pm166$    &  $(5.00\pm1.40)\times10^{-9}$   &  $66\pm18$       &  $-1.43\pm0.02$    \\
090418A  &  $1.608$   &  $1567\pm384$    &  $(4.23\pm0.26)\times10^{-10}$  &  $553\pm64$      &  $-1.63\pm0.06$    \\
090423   &  $8.2$     &  $437\pm55$     &  $(6.01\pm0.75)\times10^{-12}$  &  $376\pm115$     &  $-1.19\pm0.11$   \\
090424   &  $0.544$   &  $273\pm5$       &  $(4.32\pm0.27)\times10^{-9}$   &  $204\pm16$      &  $-1.20\pm0.01$    \\
090516   &  $4.109$   &  $971\pm390$     &  $(4.10\pm1.40)\times10^{-11}$  &  $761\pm382$     &  $-1.38\pm0.12$    \\
090618   &  $0.54$    &  $257\pm41$      &  $(1.22\pm0.03)\times10^{-9}$   &  $1206\pm47$     &  $-1.46\pm0.01$   \\
091018   &  $0.971$   &  $55\pm20$       &  $(8.25\pm0.56)\times10^{-10}$  &  $169\pm20$      &  $-1.28\pm0.02$  \\
091020   &  $1.71$    &  $280\pm190$     &  $(3.94\pm0.33)\times10^{-10}$  &  $311\pm36$      &  $-1.36\pm0.03$    \\
091029   &  $2.752$   &  $230\pm66$      &  $(1.39\pm0.08)\times10^{-11}$  &  $2806\pm343$    &  $-1.31\pm0.04$    \\
091127   &  $0.49$    &  $53.6\pm3.0$    &  $(1.03\pm0.07)\times10^{-9}$   &  $1966\pm165$    &  $-1.51\pm0.02$    \\
100621A  &  $0.542$   &  $146\pm23$      &  $(1.81\pm0.22)\times10^{-10}$  &  $3613\pm746$    &  $-1.05\pm0.03$   \\
100814A  &  $1.44$    &  $259\pm34$      &  $(1.72\pm0.08)\times10^{-11}$  &  $36217\pm4018$  &  $-2.00\pm0.08$    \\
100906A  &  $1.727$   &  $289\pm46$      &  $(1.25\pm0.17)\times10^{-10}$  &  $2341\pm402$    &  $-2.19\pm0.11$    \\
110213A  &  $1.46$    &  $231\pm21$      &  $(4.09\pm0.31)\times10^{-10}$  &  $2037\pm209$    &  $-2.13\pm0.07$    \\
110422A  &  $1.77$    &  $421\pm14$      &  $(8.40\pm0.98)\times10^{-10}$  &  $358\pm56$      &  $-1.41\pm0.03$    \\
111228A  &  $0.714$   &  $58.3\pm5.1$    &  $(7.66\pm0.53)\times10^{-11}$  &  $4094\pm464$    &  $-1.36\pm0.03$    \\
120119A  &  $1.728$   &  $515\pm22$      &  $(2.00\pm0.28)\times10^{-11}$  &  $8597\pm2226$   &  $-2.57\pm0.26$    \\
120811C  &  $2.671$   &  $198\pm11$      &  $(7.20\pm0.63)\times10^{-11}$  &  $1204\pm340$    &  $-1.57\pm0.16$  \\
120907A  &  $0.97$    &  $303\pm65$      &  $(7.26\pm0.88)\times10^{-11}$  &  $481\pm98$      &  $-1.13\pm0.04$   \\
120922A  &  $3.1$     &  $303\pm14$      &  $(1.82\pm0.67)\times10^{-10}$  &  $119\pm56$      &  $-1.10\pm0.04$   \\
121128A  &  $2.2$     &  $198\pm15$      &  $(2.61\pm0.59)\times10^{-10}$  &  $1244\pm363$    &  $-2.25\pm0.18$    \\
121211A  &  $1.023$   &  $194\pm26$      &  $(2.61\pm0.56)\times10^{-11}$  &  $2296\pm893$    &  $-1.23\pm0.10$    \\
130408A  &  $3.757$   &  $1294\pm190$    &  $(1.18\pm0.24)\times10^{-10}$  &  $216\pm118$     &  $-1.32\pm0.16$    \\
130420A  &  $1.297$   &  $129\pm7$       &  $(4.90\pm0.86)\times10^{-11}$  &  $571\pm158$     &  $-1.05\pm0.03$   \\
130427A  &  $0.3399$  &  $1161\pm7$      &  $(1.42\pm0.07)\times10^{-8}$   &  $996\pm61$      &  $-1.40\pm0.01$    \\
130505A  &  $2.27$    &  $2063\pm101$    &  $(4.54\pm0.44)\times10^{-10}$  &  $1667\pm182$    &  $-1.63\pm0.03$    \\
130610A  &  $2.092$       &  $912\pm133$     &  $(2.20\pm1.60)\times10^{-10}$  &  $41\pm35$       &  $-1.12\pm0.06$    \\
130612A  &  $2.006$   &  $186\pm31$      &  $(1.00\pm0.46)\times10^{-11}$  &  $602\pm468$     &  $-1.31\pm0.24$    \\
130701A  &  $1.155$   &  $192\pm9$       &  $(2.14\pm0.53)\times10^{-9}$   &  $50\pm15$       &  $-1.28\pm0.03$   \\
130702A  &  $0.145$   &  $14.9\pm2.3$    &  $(3.53\pm0.46)\times10^{-11}$  &  $61448\pm10744$ &  $-1.35\pm0.03$    \\
130831A  &  $0.4791$  &  $81\pm6$        &  $(1.87\pm0.29)\times10^{-10}$  &  $1925\pm332$    &  $-1.73\pm0.07$    \\
131030A  &  $1.295$   &  $406\pm22$      &  $(8.08\pm0.49)\times10^{-10}$  &  $370\pm41$      &  $-1.31\pm0.02$    \\
131105A  &  $1.686$   &  $548\pm83$      &  $(3.96\pm0.36)\times10^{-11}$  &  $986\pm235$     &  $-1.19\pm0.07$    \\
131117A  &  $4.042$   &  $222\pm37$      &  $(2.04\pm0.51)\times10^{-11}$  &  $291\pm115$     &  $-1.32\pm0.10$    \\
140206A  &  $2.73$    &  $448\pm22$      &  $(1.86\pm0.16)\times10^{-10}$  &  $1477\pm154$    &  $-1.52\pm0.03$    \\
140213A  &  $1.2076$  &  $177\pm4$       &  $(1.84\pm0.24)\times10^{-10}$  &  $1870\pm337$    &  $-1.31\pm0.04$    \\
\hline
\end{tabular}}
\end{table*}

\end{document}